\documentclass[runningheads]{llncs}
\usepackage[T1]{fontenc}
\usepackage{cite}
\usepackage{bbding}
\usepackage{float}
\usepackage{listings}
\usepackage{paralist}
\usepackage{graphicx}
\usepackage{draftwatermark}
\SetWatermarkText{PREPRINT}
\SetWatermarkScale{3}
\SetWatermarkLightness{0.93}
\usepackage{color}
\usepackage{hyperref}
\usepackage[capitalize]{cleveref}
\begin{document}
\title{Tutoring LLM into a Better CUDA Optimizer}
\titlerunning{Tutoring LLMs for CUDA}
\author{Matyáš~Brabec\orcidID{0009-0008-4470-2748} \and
Jiří~Klepl\orcidID{0000-0002-2231-4073} \and
Michal~Töpfer\orcidID{0000-0002-3313-1766} \and
Martin~Kruliš\orcidID{0000-0002-0985-8949}}
\authorrunning{M. Brabec et al.}
\institute{Charles University, Malostranské náměstí 25, 118 00 Praha~1, Czech Republic\\
\email{\{brabec,klepl,topfer,krulis\}@d3s.mff.cuni.cz}}
\maketitle
\begin{abstract}
Recent leaps in large language models (LLMs) caused a revolution in programming tools (like GitHub Copilot) that can help with code generation, debugging, and even performance optimization. In this paper, we focus on the capabilities of the most recent reasoning models to generate optimized CUDA code for predefined, well-known tasks. Our objective is to determine which types of code optimizations and parallel patterns the LLMs can perform by themselves and whether they can be improved by tutoring (providing more detailed hints and guidelines in the prompt). The generated solutions were evaluated both automatically (for correctness and speedup) and manually (code reviews) to provide a more detailed perspective. We also tried an interactive approach where the LLM can fix its previous mistakes within a session. The results indicate that LLMs are quite skilled coders; however, they require tutoring to reach optimized solutions provided by parallel computing experts.

\end{abstract}

\section{Introduction}

Large language models (LLMs) demonstrated impressive abilities to generate and modify code in various programming languages. This development has led to a revolution in programming tools, such as GitHub Copilot, that can help with code generation, debugging, and performance optimization. Recently, reasoning LLMs have been introduced~\cite{openai2024learning}. Building on the chain of thought~\cite{kojima2023large}, reasoning models solve the problem step by step by first reasoning about it (producing intermediate output) and then generating the final answer. This improves the results as the LLM can prepare the high-level structure before writing the code. We opted for using OpenAI o3-mini in this work as it is a state-of-the-art among reasoning LLMs at the time of writing.

While LLMs show promising capabilities in generating code~\cite{jiang2024survey}, many open questions remain regarding their capabilities in more complex domains like high-performance parallel computing. Although several papers have been published focusing on this topic~\cite{nichols2024can,godoy2024large}, their approach was mostly quantitative as they tried to map how many assignments from particular technologies (e.g., OpenMP, CUDA, or MPI) LLMs could solve. In this paper, we take a more detailed and narrow approach to study the quality of the generated code and how the quality correlates with the level of detail of the assignment specification.

It has been widely suggested that AI might replace junior-level programmers in the foreseeable future, so our experiments were designed with a similar philosophy. We selected three different CUDA assignments based on our experience and tried to guide the LLM in the right optimization decisions --- similarly to how a tutor or a teacher would guide a junior programmer or a student.

\subsection{Research questions and outline}

Our main objective is to assess the LLM capabilities in generating optimized CUDA code for well-known tasks to answer the following research questions:

\begin{description}
\item[Q1:] Can LLMs generate well-optimized CUDA code without specific directions?
\item[Q2:] Does the quality of the code (particularly the selected optimization) improve with more detailed prompts?
\item[Q3:] Can LLMs follow specific directions about how to optimize particular code?
\item[Q4:] Is interactive prompting (with tutoring) better than a single prompt with detailed specifications?
\item[Q5:] Can LLM set proper tuning parameters of an algorithm based on the reasoning about the given code and how it executes on GPU architecture?
\end{description}

\Cref{sec:assignments} presents the CUDA assignments we have selected for our experiments and their expected solutions.
Our prompting techniques and tutoring process are described in \cref{sec:tutoring}.
\Cref{sec:evaluation} summarizes the experimental evaluation and reviews of the generated code.
\Cref{sec:relwork} summarizes the related work, and \cref{sec:conclusion} concludes the paper.
In addition, we provide a GitHub repository\footnote{\url{https://github.com/matyas-brabec/2025-europar-llm}} where we publish all the prompts, testing scenarios, generated results, evaluation frameworks and results, and helper scripts used in our research.

\section{Assignments}\label{sec:assignments}

We chose three well-known assignments for our experiments, so the LLM should have no trouble designing correct solutions. However, they are not as profound as, for instance, matrix multiplication, so the LLM is expected to deduce the optimizations. Furthermore, we have vast experience in these assignments, which allows us to design good tutoring strategies.

\subsection{Histogram}\label{sec:assignments-histogram}

Computing histograms is a common task in data analysis. A histogram represents the number of occurrences of each value in a dataset. More specifically, we focus on a histogram that counts characters in plain text. As a minor tweak, only a (continuous) sub-range of the ASCII set is collected; other characters are ignored.

The algorithm is given an array of characters (a loaded text file) and a range of char values (\emph{from}, \emph{to}). The result is a histogram of $\emph{to}-\emph{from}+1$ bins, where bin $i$ corresponds to a character with ASCII value $i+\emph{from}$. Let us also point out that this assignment is a part of our advanced parallel programming curriculum, and it is used to introduce common CUDA optimizations to the students. Hence, we have vast experience in how to tutor students to achieve the optimal solution.

\subsubsection{Expected solutions and optimizations:}

The most straightforward solution (\textbf{baseline}) is to process one input character per CUDA thread, check that it is in the \emph{from}--\emph{to} range, and increment its corresponding bin using an atomic instruction. Both the input and the histogram are in the global memory. This solution is simple and correct but not very efficient. Since each input character can initiate an atomic write to global memory, having thousands of threads running in parallel with fewer than $256$ bins will cause the atomic operations to collide frequently.

The first optimization uses \textbf{shared memory} to store a local copy of the histogram. This way, a thread block can aggregate the updates while reducing atomic collisions since only the block updates its local copy. However, the shared memory must be initialized to zeros, and the local histogram must be copied to the global memory at the end of the thread block computation.

The effect of shared memory privatization increases with the number of input data processed by one block. Hence, a second optimization is explicitly assigning \textbf{multiple characters per thread}.

The \textbf{final} optimization further reduces the number of atomic collisions within a thread block by placing multiple histogram copies in the shared memory. We use one copy per warp lane (thread $t$ uses copy $t \bmod 32$), and each copy must be placed in its own memory bank to avoid bank conflicts. That is achieved by stridden indexing where the value $i$ of a copy $c$ is stored at offset $i \cdot 32 + c$.

Let us state that the final optimization exhibits a speedup exceeding two orders of magnitude over the baseline (using a modern Nvidia GPU and having a sufficiently large input).

\subsection{Game of Life stencil}\label{sec:assignments-gol}

The Game of Life~\cite{conway1970game} (GoL) is a cellular automaton played on a two-dimensional grid where each cell is either \emph{alive} or \emph{dead}. The assignment is to compute one iteration of the Game of Life. Given an input grid, the algorithm produces a new grid where each cell state is updated based on its eight immediate neighbors. A live cell survives only if it has two or three live neighbors; otherwise, it dies due to underpopulation or overpopulation. Meanwhile, a dead cell becomes alive if it has exactly three live neighbors, simulating reproduction.

\subsubsection{Expected solutions and optimizations:}

The \textbf{baseline} is a direct implementation of the GoL rules: each thread reads the state of one cell from global memory, examines its eight neighbors, and writes the updated state back to global memory.

One common optimization is using \textbf{shared memory} to cache a tile of cells with the corresponding halo region. However, this can improve performance only if multiple iterations are computed, which is not the case here.

A more effective strategy leverages the fact that the cell state is a boolean. Hence, one can use a bit-packed encoding using one bit per cell rather than one byte. This significantly reduces the memory footprint and improves throughput. We explored two packing strategies: \textbf{row encoding} (64-bit word represents 64 consecutive cells in a row) and \textbf{tile encoding} (64-bit word encodes an 8$\times$8 tile of cells).

Row encoding is easier to implement, but each cell has neighbors spanning over at least three different words. The tile encoding better preserves spatial locality and can better leverage instructions like \texttt{popc} to quickly compute the number of set bits (live neighbors).

Typically, one CUDA thread is assigned to compute one word. A na\"{\i}ve approach is to mask each neighboring cell separately and count them individually. This can be optimized using \texttt{popc}, which efficiently counts the number of set bits in either encoding.

The most advanced optimization implements a \textbf{full-adder}~\cite{fujita2016fast}, a 4-bit summation in a vectorized manner using bitwise instructions.
The full-adder is approximately 50$\times$ faster than the baseline.

\subsection{Nearest neighbors (kNN)}\label{sec:assignments-knn}

The $k$-nearest neighbors (kNN) algorithm is well-known in machine learning and data analysis. Its multi-query version takes $N$ data points and $M$ query points in a $d$-dimensional space, parameter $k$ (the number of neighbors to find), and a distance function. The algorithm returns indices and distances of the $k$ nearest neighbors for each query point.

To keep the difficulty of this assignment comparable to the other two, we restrict dimension $d=2$ and limit $k$ to powers of two ($32 \le k \le 1024$). With these restrictions, the best solution is a brute-force search for each query point, maintaining a top-$k$ list of nearest neighbors updated using the Bitonic Sort algorithm~\cite{zhang2023parallel}. Furthermore, we assume that $M$ (queries) is in thousands and $N$ (data points) is in millions. The assumptions are part of the specification.

\subsubsection{Expected solutions and optimizations:}

A \textbf{na\"{\i}ve} parallel implementation takes a straightforward approach where the kNN of each query is computed in one thread, and its top-$k$ set is a binary heap. Each thread iterates over all data points, updating its top-$k$ list accordingly. This solution is seemingly efficient since it requires no synchronization. However, for a GPU, it does not provide enough concurrency, introduces register pressure, and causes heavy branch divergence.

The goal of tutoring is to reach a simplified state-of-the-art implementation~\cite{zhang2023parallel}. This version assigns each query to a warp and maintains the top-$k$ sets in sorted arrays, each thread storing a run of $k / 32$ consecutive elements in registers. It also uses an equally sized buffer in shared memory to store candidate points (closer than the current $k$-th point) before they are added to the top-$k$ list. The algorithm loads data points in batches, filters them, and adds them to the candidate's buffer. When the buffer is full, it is sorted and merged with the top-$k$ set using a modified Bitonic Sort algorithm.

Since the $k$ is a power of two divisible by the warp size, the Bitonic Sort algorithm can effectively utilize the warp-shuffle operations for strides that span at least $k / 32$ elements. For smaller strides, the algorithm can run without any inter-thread communication.

\section{Tutoring}\label{sec:tutoring}

To answer the research questions Q1--Q3, we have designed a sequence of prompts with increasing levels of optimization hints and instructions where each prompt will be tested in a separate session. In some cases, separate prompts may be limiting since the LLM may not correct its mistakes or iteratively improve its solution. Therefore, we added interactive tutoring experiments (addressing Q4) where better-aimed feedback can be provided to the LLM\@.


\subsection{Single-response tests}\label{sec:single-response}

In the single-response tests, each prompt is executed in a new session, producing a single solution. The prompts are designed to incrementally introduce optimizations described in \cref{sec:assignments}. The first prompt contains only the specification and aims at question Q1. Comparing the answers to individual prompts will help us answer Q2. The most detailed prompts should answer Q3. Finally, each assignment has a set of hyper-parameters that must be selected correctly, which addresses Q5.

Each prompt starts with a clear description of the assignment and the desired code interface. It requests a kernel function(s) and a regular C++ function with a predefined signature that invokes the kernel with appropriate parameters. Everything else (host-device transfers, synchronization) is handled in our code.

In the system prompt, we instruct the LLM to act as an experienced CUDA programmer who optimizes the code for the latest Nvidia GPUs. We also explicitly ask the LLM to output only the source code and to place any additional explanations in the source code comments.

The \textbf{Histogram} assignment is divided into prompts \textbf{His1}--\textbf{His7}. \textbf{His1} is the assignment specification. It is also used as a prefix for all other prompts.

\begin{itemize}
\item \textbf{His2} and \textbf{His3} cover the shared memory optimization.
\item \textbf{His4} requests that multiple input characters be processed by each thread.
\item \textbf{His5}--\textbf{His7} cover the final optimization that should remove the local atomic and shared memory bank conflicts.
\end{itemize}

Let us note that a similar approach is taken in the advanced parallel programming course (taught at our university), where the students are guided to the optimal solution.

The \textbf{Game of Life} assignment is structured into prompts \textbf{GoL1}--\textbf{GoL6}, where \textbf{GoL1} is only the problem specification.

\begin{itemize}
\item \textbf{GoL2} suggests a row encoding and discourages the use of shared memory.
\item \textbf{GoL2~(tiled)} explores a tile encoding instead but is not pursued further.
\item \textbf{GoL3} instructs the LLM to process one 64-bit word per CUDA thread.
\item \textbf{GoL4} suggests utilizing the \texttt{popc} instruction to count active neighbors.
\item \textbf{GoL5} introduces the idea of vectorized adding without explicit instructions.
\item \textbf{GoL6} explicitly explains the full-adder technique.
\end{itemize}

The \textbf{kNN1}--\textbf{kNN8} prompts cover the \textbf{k-nearest neighbors}. Again, the first prompt provides the assignment specification, and the subsequent prompts use it as a prefix.

\begin{itemize}
\item \textbf{kNN2} and \textbf{kNN3} suggest computing the kNN of each query by a single warp while utilizing shared memory or warp shuffle instructions.
\item \textbf{kNN4} suggests updating the top-$k$ set with multiple candidates simultaneously and cooperatively (by the whole warp).
\item \textbf{kNN5} adds the candidate buffer description and its merging process.
\item \textbf{kNN6} describes the buffer management and candidate filtering.
\item \textbf{kNN7} introduces the Bitonic sort algorithm and how to use it.
\item \textbf{kNN8} outlines the memory layout of the top-$k$ set and specifies how the Bitonic sort should utilize warp shuffle instructions.
\end{itemize}

\subsection{Interactive tutoring}\label{sec:tutoring-dialogues}

One of the greatest drawbacks of the single-response tests is that the LLM does not get a chance to correct its mistakes or improve its solution. The optimization suggestions made in the prompts can be better targeted when specific lines of the generated code are referenced.
On the other hand, iterative prompting is difficult to reproduce. It can easily lead to a situation where the user tells the LLM to rewrite lines of the code almost letter by letter, which does not test the reasoning abilities of the LLM\@. To investigate the benefits of iterative prompting (Q4) in a controlled manner, we designed dialog scenarios for each assignment that should guide the prompting process. The method is similar to semi-structured interviews, which are used in research where feedback from people is required, but a fixed questionnaire is unsuitable.

The scenarios are structured as a sequence of milestones that should be achieved in a single UI session. Each milestone has an objective that describes the level of optimizations expected of the generated code. It also has the suggested initial prompt that should be used as the first step when achieving the milestone. Additionally, the scenarios describe expected hints that might be used if the LLM does not achieve the milestone on the initial prompt or list of possible issues that may be encountered during tutoring.

The \textbf{Histogram} scenario has four milestones that correspond directly to the progress of \textbf{His1} (no optimization hints), \textbf{His3} (shared memory), \textbf{His4} (multiple items per thread), and \textbf{His6} (avoiding bank conflicts) single-response prompts. The multiple items-per-thread optimization raises one particular question not covered by the single-response prompts. If a thread processes a continuous block of chars and the items-per-thread parameter exceeds 8, it may lead to uncoalesced memory loads. This can be fixed by adjusting the indexing, but we did not include this in the single-response prompts.

The \textbf{Game of Life} scenario has three milestones that correspond to the progress of \textbf{GoL1} (no optimization hints), \textbf{GoL2}\footnote{The tile-based bitwise encoding is only considered if suggested by the LLM.}--\textbf{GoL4} (row encoding and \texttt{popc}), and \textbf{GoL5}\&\textbf{GoL6} (vectorization that leads to full-adder) single-response prompts. Before introducing the full-adder logic (\textbf{GoL6}), we first examine if the LLM can devise a vectorized approach to compute multiple cells at once (\textbf{GoL5}).

The \textbf{kNN} scenario has five milestones that correspond to \textbf{kNN1} (no optimizations), \textbf{kNN2}\&\textbf{kNN3} (warp-wise top-$k$ representation), \textbf{kNN5}\&\textbf{kNN6} (shared memory candidates buffer), \textbf{kNN7} (Bitonic sort), and \textbf{kNN8} (warp-shuffle optimizations).

\subsection{Threats to validity}


The first threat is that the LLM results might be biased by the ambiguity of the assignment specifications. Although we are very well versed with the assignments, slight omissions in the specification are possible. To mitigate that, each specification was reviewed thoroughly by all authors. Subsequently, we used Grammarly\footnote{\url{https://app.grammarly.com/}} to check the grammar and spelling of the prompts since we are not native English speakers.

Intensive cooperation with LLM may lead to over-tuning the prompts to a specific model or a line of prompting that micromanages the LLM to simply rewrite the text into code. To mitigate this, we designed the prompts and scenarios strictly before the experimental evaluation.

Any LLM-based experiments may be difficult to reproduce precisely since the models use some form of randomness in the inputs (e.g., the temperature parameter). We decided to go with the default settings for the OpenAI \emph{o3-mini} model. The scripts for the API calls are available in the attached git package, and the single-response tests are repeated ten times to better understand the stability of LLM responses.

\section{Evaluation}\label{sec:evaluation}


For the single-response tests, we used the OpenAI \textbf{o3-mini} LLM with reasoning effort set to \textbf{high}. As the input for the LLM, we concatenated the system prompt and the assignment-specific prompt (as described in \cref{sec:single-response}). We used the OpenAI API\footnote{https://openai.com/api/} to obtain the LLM responses (the generated source code). Each request was repeated $10$ times. Obtaining each LLM response took around $1$--$2$ minutes. The length of reasoning varied from assignment to assignment, from 2,048--9,664 tokens for histogram up to 10,624--36,032 tokens for kNN\@.

The interactive tutoring was conducted using the OpenAI web browser UI (using the \textbf{o3-mini-high} model), and the operator followed the prescribed scenario. Each generated solution was duly evaluated before the operator proceeded with the next prompt.

The generated source codes were evaluated on three Nvidia GPU platforms: V100 (Volta), A100 (Ampere), and H100 (Hopper). In this paper, we present only the results from H100, which is the newest architecture available to us.

Complete results (LLM outputs, session histories, code reviews, and measured times for all platforms) are available in the associated GitHub repository\footnote{\url{https://github.com/matyas-brabec/2025-europar-llm}}.

\subsection{Performance tests}

\begin{figure}[ht]
  \vspace{-1.5em}
  \centering
  \begin{minipage}[b]{0.5\textwidth}
    \centering
    \includegraphics[width=\textwidth]{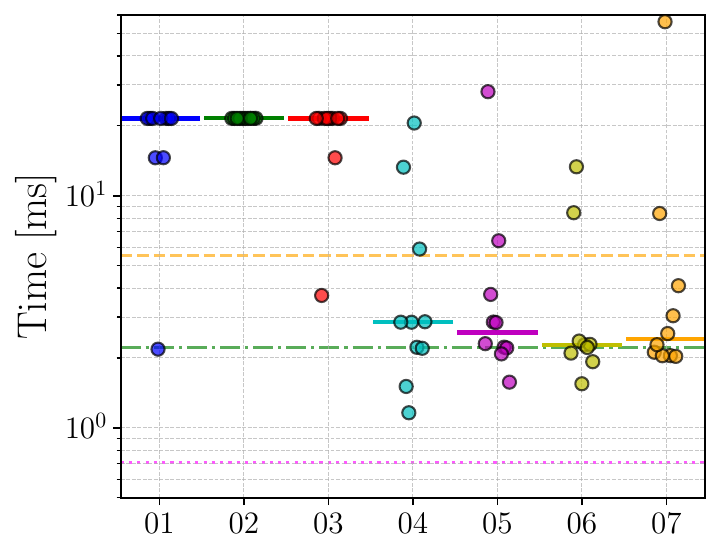}%
    \label{fig:histogram-loremipsum-perf-graph}
  \end{minipage}%
  \begin{minipage}[b]{0.5\textwidth}
    \centering
    \includegraphics[width=\textwidth]{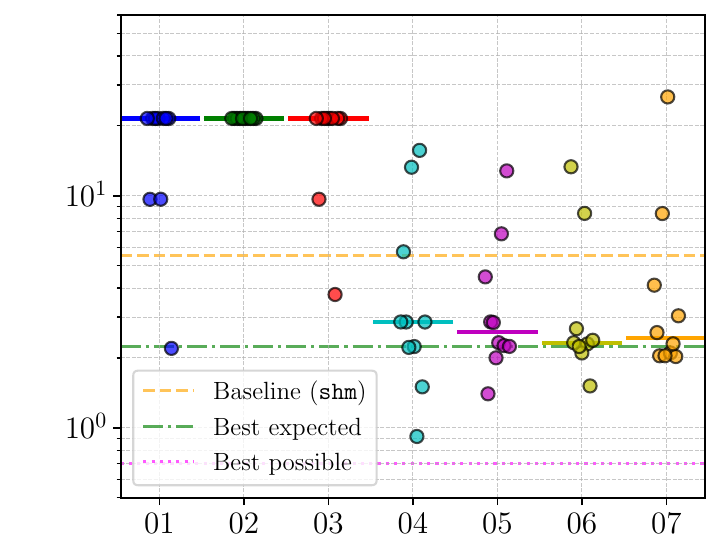}%
    \label{fig:histogram-hexdump-perf-graph}
  \end{minipage}%
  \vspace{-1em}
  \caption{Performance of \textbf{His1--7} on Lorem-ipsum (left) and Hexdump (right) inputs}%
  \label{fig:histogram-perf-graphs}
  \vspace{-1.5em}
\end{figure}

\Cref{fig:histogram-perf-graphs} shows the performance results of the histogram assignment on two inputs: \emph{Lorem-ipsum} is a randomly generated Latin-like text; \emph{Hexdump} is the hex dump of Lorem-ipsum input (i.e., hex digits and whitespace). Both inputs are repeated so that exactly 1GiB string is processed and the histogram range is set to 32--127 (printable characters). The reference measurements comprise the \emph{baseline} (shared mem. optimization), \emph{best expected} (final optimizations from \textbf{His7}), and \emph{best possible} (adding better loading pattern and tuned hyper-parameters).

Initial prompts achieved similar performance, about $3\times$ slower than the baseline, even though they use the same algorithm. The reason is that LLM insists on using 256 threads per block even though 1024 is optimal here. The graph indicates how performance improves with the tutoring. Notably, we observe a gap between the 3\textsuperscript{rd} and 4\textsuperscript{th} prompts where multiple inputs per thread are introduced. The improvements made by privatization and bank-conflicts prevention in shared memory are also distinguishable. The differences in performance (even for the same prompt) are mainly caused by selecting different items-per-thread values or different approaches to the iteration over the input.

The best-generated solution outperforms the best-expected solution and is only $1.66\times$ slower than the best possible solution. The main reason is that the LLM did not use the optimal hyper-parameters (especially the block size).

\begin{figure}[ht]
  \vspace{-1.5em}
  \centering
  \includegraphics[width=0.75\textwidth]{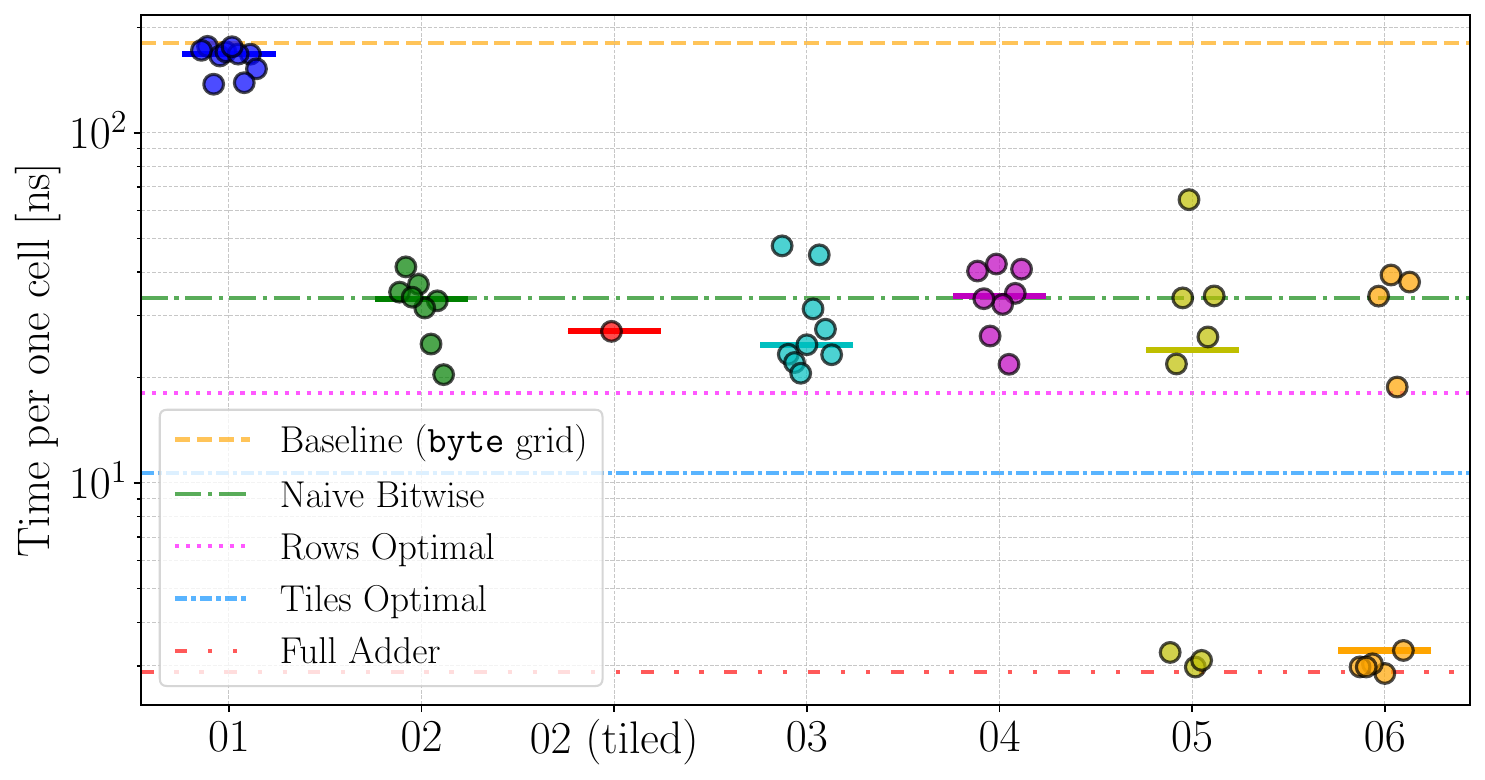}
  \vspace{-1em}
  \caption{Performance of \textbf{GoL1--6} on a 16,384$\times$16,384 grid over 200 iterations.}%
  \label{fig:gol-perf-graph}
  \vspace{-1.5em}
\end{figure}

\Cref{fig:gol-perf-graph} shows performance results for a 16,384$\times$16,384 grid over 200 iterations, normalized per iteration per cell. Five baselines were tested: \emph{Baseline} (byte grid, no optimizations), \emph{Naïve Bitwise} (individual bit masking), \emph{Rows Optimal} and \emph{Tiles Optimal} (both using \texttt{popc}), and \emph{Full Adder} (best-performing solution).

Initially, solutions failed to improve upon the baseline due to the LLM omitting bitwise encoding. \textbf{GoL2} introduced bitwise encoding, but most solutions performed similarly to the naïve approach, with only a few nearing \emph{Rows Optimal} efficiency. Tile-based solutions were largely unsuccessful, with just one functioning correctly yet underperforming compared to \emph{Tiles Optimal}. After \textbf{GoL3}, LLM generally yielded more efficient implementations, even if all suggestions have already been implemented in \textbf{GoL2}. \textbf{GoL4} was disappointing, as no solutions leveraged \texttt{popc} effectively, and median performance even declined, suggesting the LLM struggled with its correct application.

\textbf{GoL5} was the biggest surprise. Without explicit guidance, solutions varied widely, with some achieving optimal performance. Finally, \textbf{GoL6}, which explicitly instructed \emph{Full Adder} logic, produced consistently high-performing solutions.

Since the LLM was unsuccessful in generating working solutions for the majority of the kNN prompts, the performance of the generated solutions is not presented. A more detailed analysis of the generated solutions is included in \cref{sec:knn-discussion}.

\subsection{Code reviews and discussion}

The \textbf{histogram} was solved quite well by the LLM since it is the simplest of the selected assignments. All generated solutions were compilable (some required minor correction), and only one solution (out of 70) failed (due to an indexing error in the code).

The first suggested optimization (shared memory) was employed in all solutions even when it was not specifically requested since it is a textbook optimization demonstrated in many tutorials. All the solutions also prepared the kernel to be able to handle multiple input chars per thread; however, only one solution (from \textbf{His1--3}) used the correct parameters to exploit it. Unexpectedly, the LLM created a minor optimization, adding an \texttt{if}-statement to skip the atomic add for bins that remain zero (when the local copy is merged into the global memory). This was not suggested in the prompt since it works only in special cases where shared memory is used, but not enough values are aggregated there.

All solutions past the \textbf{His4} implemented the multiple items per thread optimization. They used one of two trivial approaches --- each thread processes a continuous block or uses block size as the loop stride. Both approaches are suboptimal; however, the optimal pattern was not specified in the prompts and was not expected in the solutions.

The final optimization (multiple histogram copies in the sharded memory) was properly implemented only for the \textbf{His7} prompt, where the proper layout is explicitly expressed as an equation. Prompts \textbf{His5} and \textbf{His6} led to solutions where multiple copies were present, but their memory layout caused bank conflicts. Some solutions tried alternative approaches, such as histogram padding or placing a private copy per thread; however, these approaches do not work.

\subsubsection{Game of Life:}

Across all prompts, the LLM produced syntactically correct and mostly functional code. However, \textbf{GoL2~(tiled)} proved particularly challenging, with only one correct solution due to the complexity of handling edge cases in a tiled layout.

The LLM consistently applied textbook CUDA optimizations, such as shared memory (\textbf{GoL1}) and bitwise operations (\textbf{GoL2}--\textbf{GoL3}). However, advanced techniques like lookup tables, tile-based bit packing, and \texttt{popc} were only used when explicitly prompted. Even then, results were mixed: in \textbf{GoL4}, most implementations misused \texttt{popc}, failing to leverage its performance benefits. Surprisingly, the LLM successfully applied full-adder logic in \textbf{GoL5}—despite no explicit hint—likely due to its prevalence in existing implementations.

Highly specific prompts (\textbf{GoL6}, full-adder technique) led to correct and consistent solutions, while open-ended ones (\textbf{GoL5}) encouraged creativity, producing unexpected methods such as vector-like word processing and large lookup tables. However, these novel approaches often introduced inefficiencies or outright errors, such as incorrect indexing in a lookup table (\textbf{GoL4}) and frequent edge-case failures in the tiled implementation (\textbf{GoL2~(tiled)}).

The LLM effectively generated correct CUDA kernels for GoL when given structured hints. Open-ended prompts led to more diverse solutions, however, at the cost of correctness and efficiency. This reinforces the trade-off between guidance and innovation: precise instructions improve reliability, while ambiguity fosters creativity—sometimes at the expense of performance.


\subsubsection{kNN:}\label{sec:knn-discussion}

For this assignment, the LLM showed a mediocre programming competence. A drop in performance was expected since the kNN algorithm consists of more individual steps than Histogram or Game of Life assignments. In the main loop, we need to load the data points, compute the distance to the query point, and insert the data point into the top-$k$ list. With the expected optimizations, the programming complexity increases even further.

Even for the first prompt (\textbf{kNN1}), which puts no restrictions on the solution design, only half of the generated solutions provided a valid result for the specified input parameters. The common characteristic of the successful solutions was the lack of inter-thread communication. Most of these solutions were similar to the na\"{\i}ve solution. For the other prompts (starting with \textbf{kNN2}), the LLM was unable to provide a correct solution, and most of the solutions either crashed during execution or required a forceful termination due to a deadlock.

From the first prompt onward, a recurring pattern emerged: the LLM attempts to distribute each query's intermediate top-$k$ result among warp threads, each storing a $k / 32$-sized portion of the array locally. This approach is explicitly specified only in the last prompt, as it requires a careful algorithm design. In the generated solutions using this approach, each thread processes different data points and enqueues them into its $k / 32$-sized list without any inter-thread communication. Since the LLM does not employ inter-thread communication to either share the processed data points or to combine the local lists into the intended top-$k$ list, the algorithm effectively becomes a parallelized $k / 32$-nearest neighbor search, and the final result is thus incorrect. This issue was present in virtually all solutions that used this approach; this approach was used in the vast majority of the solutions for all prompts, starting with \textbf{kNN3}.

The solutions showed a range of designs, storing the intermediate results in priority queues represented as binary heaps, sorted arrays, or unsorted arrays. Heaps were most common in \textbf{kNN1}, unsorted arrays in \textbf{kNN2} and \textbf{kNN3}, and sorted arrays in all other prompts. Binary heaps and sorted arrays both allow for efficient pre-filtering, requiring only one comparison. Binary heaps offer the most efficient insertion while updating the unsorted arrays requires only a single write operation. However, in a parallelized approach, the best performance is achieved by using sorted arrays as they allow for the use of Bitonic sort for efficient merging with a buffer of new elements. Generally, the LLM was able to correctly manage any of the three data structures.

The biggest struggle for the LLM was inter-thread communication and the semantics of the code in the context of CUDA parallelism. In this regard, the LLM often made the two following mistakes: \begin{inparaenum}[(i)]
\item The algorithm selects a specific lane, which then performs a parallel algorithm with inactive threads (e.g., performing warp shuffle operations or doing warp-wide synchronization).
\item The LLM inconsistently uses some variables as local or shared --- for example, it declares a variable as local but then updates it by selecting a specific lane.
\end{inparaenum}

In conclusion, the kNN assignment proves that the LLM is not sufficiently capable of reasoning about the CUDA parallelism if it requires inter-thread communication and synchronization.

\subsection{Interactive tutoring}\label{sec:evaluation-interactive}

The interactive tutoring did not generate completely different versions of the code. The main difference was that the LLM was more reluctant to change the decisions made in the previous prompts. From this perspective, the repeated single-response prompts provided more variety (especially in the hyperparameter selection) that could be perceived as a simplified educated autotuning process.

The testing of the \textbf{histogram} scenario went smoothly; the LLM even solved the second milestone within the first one. We decided to take the testing slightly further than the scenario specified, and the result got closer to the best possible solution when extra instructions on iterating over the input were provided. However, these instructions needed to be very specific to be understood by the LLM\@.

The \textbf{Game of Life} scenario proceeded as expected but with some inconsistencies. The LLM correctly implemented the base solution and transitioned to a bitwise representation. However, it misused \texttt{popc}, and when tasked with updating multiple cells at once, it failed to vectorize and instead reused \texttt{popc}. After explicitly hinting at the full-adder technique, it produced a flawed vectorized version but later corrected its mistake. Still, the final solution remained one of the least efficient in the \textbf{GoL6} group.

The interactive tutoring was relatively unsuccessful for \textbf{kNN}, just like the single-response prompting. The LLM performed better as it was able to fix its previous mistakes and use the simpler solutions as a base for the next iteration. However, it made the very same mistakes in the context of inter-thread communication. In milestone 4, which relies on it, the LLM could not fix its mistakes without introducing new ones or repeating the previous ones, and the process became unproductive.

\subsection{Answering research questions}
\begin{description}
\item[Q1:] Yes, simple tasks can be optimized quite well by the LLM without specific instructions; however, more complex solutions require tutoring.
\item[Q2:] Yes, the quality of the code improves with the level of detail in the prompt.
\item[Q3:] Yes, we have observed that the LLM was able to apply even very specific instructions to improve the code.
\item[Q4:] Iterative prompting may help with fixing errors or adjusting parameters; however, LLM also tends to keep the previous decisions (like the hyperparameter selection) unless it is told to change them, which limits the exploration of parameter space or possible decisions.
\item[Q5:] LLM seems to be very conservative regarding the proper hyperparameter selection (e.g., block size). It might be beneficial to include autotuning tools in the code design process or to provide performance feedback to the LLM.
\end{description}

\section{Related Work}\label{sec:relwork}

Since LLMs have shown impressive capabilities in general programming tasks, there has been a steep increase in interest in using these models for parallel programming.
Nichols~et~al.~\cite{nichols2024can} constructed ParEval, a benchmark for evaluating LLMs on parallel programming tasks. They conclude that LLMs are significantly better at generating serial code than parallel code and that the generated parallel code often fails to perform and scale as expected. They also evaluate the performance of LLMs on stencil and histogram algorithms in CUDA\@. For these tasks, the authors show that the (best performing) GPT-4 model is capable of generating parallel code with 80\% and 58\% pass rates, respectively. However, for search problems, the GPT-4 model only achieves a pass rate of 27\%. This is consistent with our findings that the kNN algorithm is more difficult for LLMs to generate. Furthermore, the authors show that, without sufficient guidance, the LLMs do not generate efficient parallel code.

A specific approach to CUDA code generation is presented by Palkowski et~al.~\cite{palkowski2024gpt} on Nussinov's algorithm for RNA folding prediction. They propose a multi-stage approach that uses a polyhedral compiler to generate OpenMP-parallelized code and then translates the OpenMP code to functionally equivalent CUDA code. They show that, with this approach, GPT-3.5 provides a parallel code with satisfactory performance; however, they also conclude that it is unreliable in following code generation guidelines.

Chen~et~al.~\cite{chen2023vscuda} evaluate the performance of LLMs in generating CUDA code in an interactive environment. They show that the GPT-4 model can apply specific optimizations to CUDA code, especially when re-prompted (similar to iterative tutoring). Godoy~et~al.~\cite{godoy2024large} also explore interactive code generation with LLMs and show that GPT-3.5, a slightly older model, is unreliable in auto-parallelization tasks involving generating CUDA-parallelized code from serial base code without specific guidance.

To the best of our knowledge, our work is the first to study the applicability and effects of tutoring to LLM-based CUDA kernel generation using reasoning LLMs, which often outperform the non-reasoning models~\cite{kojima2023large}. The related works show that even non-reasoning LLMs can generate CUDA code, but their performance can be inconsistent. Since the works show that the efficiency of the generated code without any guidance is often unsatisfactory, our work provides an important step towards improving the performance of LLMs as we show that the tutoring approach improves the reliability and the performance of the CUDA code generated by reasoning LLMs.

\section{Conclusion}\label{sec:conclusion}

In this work, we have used three well-known CUDA assignments to evaluate the capabilities of reasoning LLMs in generating optimized CUDA code. We have shown that the LLMs can generate correct solutions for the assignments, but they often lack the optimizations that are crucial for performance. The quality of the solution can be improved by tutoring, where LLM is given more detailed specifications and suggestions about possible optimizations. The simple optimizations can be applied by LLM itself just based on an appropriate suggestion. More complex optimizations need to be detailed at the level of algorithm descriptions or equations. In the case of more complex assignments (kNN), the LLM often fails to generate correct solutions without tutoring.

In summary, the LLMs are likely to perform a similar role as junior-level programmers. They are quite good at following instructions, but they can rarely make the right higher-level decisions without appropriate guidance. Furthermore, the models often fail to select optimal hyperparameters for an algorithm, so it is still crucial to make performance evaluations or even autotuning along with LLM code generation. In the future, we plan to investigate how to use evaluation results as better feedback for the LLMs.

\begin{credits}

\subsubsection{\ackname}
This work was partially supported by the Johannes Amos Comenius Programme (P JAC) project CZ.02.01.01/00/22\_008/0004605 (Natural and anthropogenic georisks),
the EU project ExtremeXP grant agreement 101093164, 
by the Charles University institutional funding 260821, 
and by the Charles University Grant Agency project 269723. 

\subsubsection{\discintname}
The authors have no competing interests to declare that are relevant to the content of this article.

\end{credits}

\bibliographystyle{splncs04}
\bibliography{bibliography}

\end{document}